\documentclass[showpacs,nofootinbib,preprintnumbers,amsmath,amssymb,preprint,aps]{revtex4}
\usepackage[dvips]{graphicx}
\usepackage{graphicx}
\usepackage{amsfonts}
\usepackage{bm}
\usepackage{amsmath}
\usepackage{amssymb}
\usepackage{color}
\usepackage[all]{xy}

\def\be{\begin{equation}}
\def\ee{\end{equation}}
\def\bea{\begin{eqnarray}}
\def\eea{\end{eqnarray}}

\begin{document}

\title{Einstein-Maxwell-Dilatonic phantom black holes: Thermodynamics and geometrothermodynamics}

\author{Hernando  Quevedo$^{1,2}$, Mar\'\i a N. Quevedo$^3$ and Alberto S\'anchez$^4$}
\email{quevedo@nucleares.unam.mx,maria.quevedo@unimilitar.edu.co,asanchez@nucleares.unam.mx}
\affiliation{
$^1$Instituto de Ciencias Nucleares, Universidad Nacional Aut\'onoma de M\'exico,
 AP 70543, M\'exico, DF 04510, Mexico\\
$^2$Dipartimento di Fisica and ICRANet, Universit\`a di Roma ``La Sapienza",  I-00185 Roma, Italy\\
$^3$ Departamento de Matem\'aticas, 
 Facultad de Ciencias B\'asicas, 
Universidad Militar Nueva Granada, Cra 11 No. 101-80, 
Bogot\'a D.E., Colombia \\
$^4$Departamento de Posgrado, CIIDET,  AP752, 
Quer\'etaro, QRO 76000, Mexico}

\date{\today}

\begin{abstract}

We use the Legendre invariant formalism of geometrothermodynamics to investigate the geometric properties of the equilibrium space  of a 
spherically symmetric phantom black hole with electric charge and dilaton. We find that at certain points of the equilibrium space, the thermodynamic curvature is characterized by the presence of singularities that are interpreted as phase transitions. We also investigate the phase transition structure by using the standard approach of black hole thermodynamics based upon the analysis of the heat capacity and response functions. We show compatibility between the two approaches. In addition, a new type of phase transition is found which is due to the presence of phantom energy and corresponds to a transition between black hole states with different stability properties.


\end{abstract}


\maketitle

\section{Introduction}
\label{sec:int}

Black hole thermodynamics has been the subject of intensive investigation since its formulation about forty years ago \cite{bek73,haw74,haw75,dav77}. 
The reason is that it is considered as an indication of the quantum nature of the black hole interior. Although many attempts have been made to 
find a statistical formulation of black hole thermodynamics, no definite statistical model is known today \cite{car09}. Indeed, this issue is closely 
related to the quantization of gravity, one of the major problems of modern theoretical physics. 
In an ordinary  system, thermodynamic properties are the macroscopic limit of some microscopic model, usually specified through the partition 
function. For instance, temperature is interpreted as a measure of the average energy of microscopic constituents, and entropy counts the number of microscopic states. 
The natural question arises whether the same is true for  black holes. The answer to this question might shed some light on the problem of quantum gravity, 
because the Bekenstein-Hawking entropy for black holes contains the (quantum) Planck constant and the (gravitational) Newton constant. This makes black hole thermodynamics an interesting subject of investigation. 

To investigate the thermodynamic properties of black holes, one usually starts from the fundamental equation $S=A/4$ that relates entropy $S$ with the horizon area $A$.
From a thermodynamic point of view both variables are extensive and, therefore,  the fundamental equation should be a homogeneous function of some degree \cite{dav77}.
The first law of black hole thermodynamics permit us to compute all the corresponding intensive variables, and to perform the analysis of the temperature behavior, stability, phase transitions, etc. This approach can be considered physical in the sense that it is based upon the assumption of the validity of the laws of classical thermodynamics.

On the other hand, the properties of a thermodynamic system can also be investigated by using the formalism of thermodynamic geometry which consists in equipping the 
space of equilibrium states with a Riemannian geometric structure.  This idea was first implemented in statistical physics and thermodynamics
by Rao \cite{rao45}, in 1945, by introducing a metric whose components in local coordinates coincide with Fisher's
information matrix. Rao's method has been applied and generalized  by a number of
authors (see, e.g., \cite{amari85} for a review). Moreover, Riemannian geometry in the space of
equilibrium states was introduced by Weinhold \cite{wei75} and Ruppeiner \cite{rup79,rup95},
who defined metric structures as the Hessian of the internal energy and (negative of) the entropy, respectively.
Both metrics have been used intensively to study the geometry of the thermodynamics of
ordinary systems and black holes (see, for instance, \cite{rev} for a review). 

 Recently, an alternative mathematical approach called geometrothermodynamics (GTD) was proposed in \cite{quev07}. It is based upon the 
assumption that all the geometric objects that enter the formalism should be invariant with respect to Legendre transformations, which in classical thermodynamics corresponds to the well-known fact that the properties of a system do not depend on the choice of thermodynamic potential \cite{callen}. 
GTD uses the geometric properties of the equilibrium space to describe the thermodynamic properties of the corresponding system. For instance, since the equilibrium space is endowed with a Riemannian metric, 
the Riemann curvature tensor is interpreted as a measure of thermodynamic interaction, and the curvature singularities correspond to phase transitions. GTD has been shown to be true in a large number of black hole configurations \cite{appl}. 
However, in the case of phantom black holes \cite{jrh12,ro12}, it seems to lead 
to contradictions. Indeed, Rodrigues and Oporto recently investigated in the framework of GTD  a class of spherically symmetric black holes with phantom charge and dilaton, and found that curvature singularities exist in the equilibrium space which do not correspond to divergencies of the heat capacity, i.e., they cannot be identified as phase transitions \cite{ro12}. Of course, one could argue that phantom black holes show a pathological behavior due to fact that they are characterized by negative energy densities and, therefore, the formalism of GTD leads to inconsistencies in the case of pathological configurations. Nevertheless, the point is that black hole thermodynamics can handle even such pathological situations and does not lead to inconsistencies, although the thermodynamic behavior is not quite physical. 
So, in principle, one should demand that GTD should also be able to handle such pathological configurations. 

On the other hand, there is also a physical argument 
in favor of the existence of phantom fields in nature. The recently observed acceleration of our Universe suggests the existence of an exotic fluid with negative pressure that is the source of the repulsive gravitational force necessary to generate acceleration. However, repulsive gravity can also be generated by 
fields with negative energy density. In fact, observational data \cite{hann06,dun09} suggests that a phantom field could also explain the acceleration of our Universe.  

The purpose of the present work is to show that GTD is able to correctly describe the thermodynamics of the phantom dilatonic black holes presented in \cite{ro12}. 
In fact, we will show that the curvature of the equilibrium space predicts the existence of three types of phase transitions. The first one corresponds to a divergence of the heat capacity of the black hole. The second one corresponds to the divergence of a particular response function and, therefore, it is corroborated by classical black hole thermodynamics. The third one occurs when the capacity and all the response functions vanish, indicating a drastic change between states with different stability properties. We argue that the appearance of this third type of singularity is due to the exotic nature of the matter that generates the black hole.

This work is organized as follows. In Sec. \ref{sec:pdbh}, we present the explicit form of the black hole and discuss its fundamental equation and the main thermodynamic properties. Then, in Sec. \ref{sec:pts}, we derive the Legendre invariant metric for the equilibrium space  and compute the corresponding
thermodynamic curvature to find the phase transition structure of this class of phantom black holes. We show that our results predict phase transitions that 
can be corroborated by the behavior of the heat capacity and response functions of the black hole, according to classical black hole thermodynamics. 
In Sec. \ref{sec:con}, we discuss our results. Throughout this paper we use geometric units with $G=c=\hbar = k_{_B}=1$.


\section{Phantom dilatonic black holes}
\label{sec:pdbh}

The Einstein-Maxwell Lagrangian density with a dilaton field can be expressed as (we follow here the notations and conventions of Ref. \cite{ro12}; see also 
\cite{nrm12,nrn15})
\be
{\cal L} =
R - 2 \eta_1\, g^{\mu\nu} \varphi_{,\mu} \varphi_{,\nu} + \eta_2\, e^{2\lambda \varphi}F_{\mu\nu} F^{\mu\nu} \ ,
\label{action}
\ee
where $R$ is the scalar curvature, $F_{\mu\nu}$ is the Faraday tensor of the electromagnetic field, and $\varphi$ represents the dilaton field.
The non-minimal coupling between the electromagnetic and dilatonic fields is represented by the real constant $\lambda$. 
The two parameters $\eta_1$ and $\eta_2$ can be so chosen that they determine the nature of the corresponding fields. So, $\eta_2=+1$ represents 
the classical Maxwell field whereas $\eta_2=-1$ indicates that the contribution of the electromagnetic energy to the action is negative, which is the reason
why it is called phantom. Moreover, for $\eta_1=-1$ the dilatonic field is phantom and for $\eta_1=+1$ it represents the classical dilaton.
 The constant $\lambda$  determines the special theories contained in Eq. (\ref{action}). 
 For $\lambda = \sqrt{3}$,  the Lagrangian 
(\ref{action}) leads to the Kaluza-Klein field equations obtained from
the dimensional reduction of the five-dimensional Einstein vacuum equations. For $\lambda =  1$,
the Lagrangian coincides with the low energy limit of string theory with vanishing dilaton
potential. Finally, in the extreme limit $\lambda =   0$ , Eq. (\ref{action}) reduces to the Einstein–Maxwell
theory minimally coupled to the scalar field. The structure of the field equations and some particular classes of solutions have been 
investigated in \cite{mnq95}.

For the theory following from the action (\ref{action}) a particular spherically symmetric solution was derived in \cite{cfr09} which is represented by the line element
\be
ds^2 = f_1(r) dt^2 - \frac{dr^2}{f_1(r)} - r^2 f_2(r) (d\theta^2 + \sin^2\theta d\phi^2) \ ,
\label{lel}
\ee
and by the electric and dilatonic fields
\be
F= \frac{1}{2} F_{\mu\nu} dx^\mu\wedge dx^\nu = \frac{q}{r^2} dt\wedge dr  \ , \quad e^{-2\lambda\varphi} = f_2(r) \ ,
\ee
where
\bea 
f_1(r) = & \left(1-\frac{r_+}{r}\right)\left(1-\frac{r_-}{r}\right)^\gamma \ , \\
f_2(r) = & \left(1-\frac{r_-}{r}\right)^{1-\gamma} \ .
\eea
The constant $\gamma$ is defined as
\be
\gamma=\frac{1-\eta_1\lambda^2}{1+\eta_1\lambda^2}= \frac{\lambda_-}{\lambda_+} \in \Bigg\{ 
\begin{array}{ll} 
                                       (-1,1) & \quad {\rm for}\quad \eta_1 = 1 \\
                                      (-\infty,-1) \cup (1,+\infty) & \quad {\rm for}\quad \eta_1 = - 1 \ 
\end{array}		
\label{gamma}										
\ee
Moreover, the constants $r_\pm$ are given in terms of the mass $M$ and charge $q$ as 
\bea
r_+ = & M + \sqrt{M^2 -\frac{2\eta_2\gamma q^2 }{1+\gamma} } \ , \\
r_- = & \frac{1}{\gamma}\left( M - \sqrt{M^2 -\frac{2\eta_2\gamma q^2 }{1+\gamma} } \right) \ ,
\eea
and are subject to the conditions 
\bea
0< r_- < r_+ & \quad  {\rm for}\quad \eta_2\lambda_+ > 0\ , \label{cond1} \\
r_- < 0<  r_+ & \quad {\rm for}\quad \eta_2\lambda_+ < 0\ . \label{cond2}
\eea
This is a black solution with an inner horizon located at $r=r_-$ and an outer event horizon at $r=r_+$. 

The fundamental thermodynamic equation, $S=A/4 = (1/4) \int  \sqrt{g_{\theta\theta}g_{\phi\phi} } d\theta d\phi$, can be calculated explicitly by using the line element (\ref{lel}), 
\be
S = \pi r_+^{1+\gamma} (r_+ - r_-)^{1-\gamma} \ .
\label{ent}
\ee
Notice that the entropy is given in units of $lenght^2$, as expected in geometric units. The entropy 
must satisfy the first law of black hole thermodynamics \cite{ro12}
\be
dS = \frac{1}{T} dM - \eta_2 \frac{A_0}{T} d q \ ,
\ee
where $T$ is the temperature and $A_0$ the electric potential at the horizon. 
Then, a straightforward computation shows that
\be
T= \frac{(r_+-r_-)^\gamma}{4\pi r_+^{1+\gamma} }\ , \quad A_0  = \frac{q}{r_+}\ .
\ee

Notice that $r_+$ and $r_-$ are first-degree homogeneous functions of the extensive variables $M$ and $q$. This implies that the entropy (\ref{ent}) is a 
second-degree homogeneous function. The mass variable $M$ cannot be found explicitly from Eq.(\ref{ent}) in terms of $S$ and $q$ because that function is not invertible. However, it can be expressed in terms
of the horizons radii as 
\be
M= \frac{1}{2} (r_+ + \gamma r_-)\ .
\ee
We see that all thermodynamic variables are well-behaved functions of $r_+$ and $r_-$. This means that in terms of $M$ and $q$, the entropy and temperature do not 
present any peculiar behavior as far as the radii $r_\pm$ are well-behaved functions of $M$ and $q$. However, the particular case for which  $r_+ = r_-$ leads to the  vanishing of the temperature, horizon area and entropy, indicating that the fundamental equation is not well-defined at that point. Moreover, for $r_+=r_-$ 
the mass of the black hole could also be negative for certain values of the parameter $\gamma$. Hence, we limit ourselves to the investigation of the case
\be
r_+> r_- \ .
\ee
In order for the outer horizon radius to be well-defined we also suppose that $r_+>0$.


\section{Geometrothermodynamic phase transition structure}
\label{sec:pts}

One if the most important properties of a black hole is its phase transition structure. Due to the lack of a complete microscopic model, it is still not possible to describe the physical changes that occur during a phase transition. Nevertheless, from a thermodynamic point of view, phase transitions indicate that the equilibrium properties of the system are no longer valid and, instead, we should use a different approach (maybe non-equilibrium thermodynamics) to investigate the physical processes that accompany a phase transition. In the context of GTD, a phase transition should also indicate that the equilibrium description breaks down. We expect therefore that in GTD a phase transition should correspond to a curvature singularity of the equilibrium space. In this section, we will investigate this question in the case of the phantom dilatonic black holes presented in the previous section.  

\subsection{The formalism of geometrothermodynamics}

Let us recall that one of the objectives of GTD is to construct a formalism that is invariant with respect to Legendre transformations. This is an important condition because classical thermodynamics does not depend on the choice of the thermodynamic potential, and different potentials are related by means of Legendre transformations. A more detailed explanation of Legendre transformations and Legendre invariance is given in Appendix \ref{sec:leginv}.  
In our first attempt to construct such a formalism \cite{quev07}, we first noticed that Hessian metrics, which have been used as the basis of thermodynamic geometry,
are not Legendre invariant. This means that the geometric properties of the equilibrium space can change as the thermodynamic potential is changed.
Then, we proved  that the simplest way to make a Hessian metric Legendre invariant is to ``multiply" it by the corresponding thermodynamic potential 
\cite{quev07,quev08}. In fact, we now know that this is the only way to reach Legendre invariance, if we limit ourselves to total Legendre transformations and 
impose the additional physical condition that the  curvature tensor should vanish if thermodynamic interaction is lacking \cite{gl14} (see Appendix \ref{sec:tdint}). 
Then, we noticed that 
it is necessary to choose a pseudo-Euclidean signature of the Legendre invariant metric to correctly describe black hole thermodynamics \cite{aqs08}. This is the situation so far. Now the question is how to ``multiply" by a potential the metric of the equilibrium space in a Legendre invariant way. It turns out that to handle  Legendre transformations as coordinate transformations in differential geometry, it is necessary to introduce an auxiliary space called phase space. 

To be more explicit, let us consider a contact Riemannian manifold $({\cal T},\Theta, G)$, where ${\cal T}$ is a $(2n+1)-$dimensional manifold, $\Theta$ is a contact form, i.e., it satisfies the condition $\Theta\wedge (d\Theta)^n \neq 0$ and $G$ is a Riemannian metric. If we choose the set $Z^A = \{\Phi, E^a, I^a\}$ with $A=0,1,...,2n$ and $a=1,...,n$, according to Darboux theorem, the canonical representation of the contact form is $\Theta = d\Phi - \delta_{ab} I^a dE^b$, and a Legendre transformation can be represented as the coordinate transformation \cite{arnold}
\be
\{Z^A\}\longrightarrow \{\widetilde{Z}^A\}=\{\tilde \Phi, \tilde E
^a, \tilde I ^ a\}\ , 
\ee
 \be
 \Phi = \tilde \Phi - \delta_{kl} \tilde E ^k \tilde I ^l \ ,\quad
 E^i = - \tilde I ^ {i}, \ \
E^j = \tilde E ^j,\quad
 I^{i} = \tilde E ^ i , \ \
 I^j = \tilde I ^j \ ,
 \label{leg}
\ee 
where $i\cup j$ is any disjoint decomposition of the set of
indices $\{1,...,n\}$, and $k,l= 1,...,i$. In particular, for
$i=\emptyset$ we obtain the identity transformation. Moreover, for
$i=\{1,...,n\}$, Eq.(\ref{leg}) defines a total Legendre
transformation, i.e., \be
 \Phi = \tilde \Phi - \delta_{ab} \tilde E ^a \tilde I ^b \ ,\quad
 E^a = - \tilde I ^ {a}, \ \
 I^{a} = \tilde E ^ a \ .
 \label{legtotal}
\ee 

We define the thermodynamic phase space as a Legendre invariant contact Riemannian manifold. It is easy to see that the contact form $\Theta$ is invariant with respect 
to Legendre transformations. As for the metric, the situation is more complicated. In fact, the set of Legendre transformations do not form a group and hence 
it is not possible to use the standard methods of differential geometry to generate the most general Legendre invariant metric. Nevertheless, in the case of total Legendre transformations, it is possible to  identify a quite general metric in the form \cite{qq11,qqst14}\footnote{It is still possible to multiply both terms of the metric by Legendre invariant functions (or constants)  $\Lambda_1(Z^A)$ and $\Lambda_2(Z^A)$, respectively, without affecting the main results.}
\be
G^{^{I/II}} = \Theta^2 + (\xi_{ab} I^a E^b) (\chi_{cd} dI^c dE^d) \ ,
\label{gup}
\ee
where $\xi_{ab}$ and $\chi_{ab}$ are diagonal constant $(n\times n)$-matrices. It turns out that if we choose $\chi_{ab} = \delta_{ab}= {\rm diag}(1,\cdots,1)$, the resulting metric $G^{^I}$ can be used to investigate systems with at least one first-order phase transition. Alternatively, if we choose 
$\chi_{ab} = \eta_{ab}= {\rm diag}(-1,\cdots,1)$, we obtain a metric $G^{^{II}}$ that correctly describes systems with second-order phase transitions. 
This is the case of black holes. 

In GTD, a thermodynamic system is described by its corresponding equilibrium space ${\cal E}$  which is defined as follows. Let 
$\varphi: {\cal E} \rightarrow {\cal T}$ or, in coordinates, $\varphi:\{E^a\}\mapsto \{\Phi(E^a), E^a, I^b(E^a)\}$,
  be a smooth embedding  map which satisfies the condition 
$\varphi^*(\Theta) =0$, i.e., $d\Phi = \delta_{ab}I^a dE^b$ and, consequently, 
 $\partial \Phi /\partial E^a \equiv \Phi_{,a}= I_a \equiv \delta_{ab} I ^b$ on ${\cal E}$. The pullback $\varphi^*$ induces a canonical  metric $g=\varphi^*(G)$
on ${\cal E}$. For instance, (modulo an ignorable multiplicative constant)
\be
g^{^{II}}= \varphi^*(G^{^{II}}) = \Phi \, \eta^b_a \Phi_{,bc} \, d E^a dE^c\ ,
\label{gdown}
\ee
where $\eta_a^c ={\rm diag}(-1,1,\cdots,1)$. This is how the metric of the equilibrium space becomes ``multiplied by the potential" in a Legendre invariant way.
Indeed, the conformal factor $  \xi_{ab} I^a E^b$ in the metric (\ref{gup}) transforms under the pullback  as 
$\xi_{ab} E^b  \varphi^*(I^a)= \xi_{ab} E^b \delta^{ac}\Phi_{,c} \sim \Phi$. Here we have used Euler's identity in the following form. If $\Phi$ is a homogeneous 
function of degree $\beta$, i.e., $\Phi(\lambda E^a) = \lambda^\beta \Phi(E^a)$, then Euler's identity reads $E^a \Phi_{,a} =\beta \Phi$. If $\Phi$ is a generalized
homogeneous function \cite{sta71}, i.e., $\Phi(\lambda^{\alpha_a} E^a ) = \lambda \Phi(E^a)$, then $\alpha_a E^a \Phi_{,a} = \alpha_\Phi \Phi$. This issue has been discussed in detail in \cite{turco}.   
In any case, we see that the metric of the equilibrium space gets the potential $\Phi$ as conformal factor, in accordance to the Legendre invariance requirement.

When written in the form (\ref{gdown}), the thermodynamic metric  $g^{^{II}}$ of the equilibrium space does not seem to have any particular physical significance. However, it is 
possible to use the properties of the phase space generating metric (\ref{gup}) to show the explicit components of $g^{^{II}}$ have a significance in fluctuation theory. This is shown in the Appendix \ref{sec:phys}.   


\subsection{Phase transitions}

We now consider the particular case of the black hole with fundamental equation $S=S(M,q)$ given in Eq.(\ref{ent}). Accordingly, 
the thermodynamic potential is $S$ and $E^a=\{M,q\}$. Then, from Eq.(\ref{gdown}) we obtain the metric 
\be
g= \frac{-4\pi^2 r_+^{1+2\gamma}} {(r_+-\gamma r_-)(r_+-r_-)^{2\gamma}}\bigg[ 2 r_+ (r_+ -2\gamma r_- - r_-) d M^2 + [r_+^2 +\gamma(1+2\gamma)r_-^2 +(\gamma-1)r_+ r_-] dq^2 \bigg]\ ,
\label{gd}
\ee
where we dropped the index $II$ and re-expressed the final results in terms of $r_\pm$ for simplicity. Notice that the components of the metric have units of 
$length^2$.
 A lengthy computation leads to the following curvature scalar
\be
R = \frac{4 T^2 }{\gamma} \frac{N(M,q)}{D(M,q)} \ ,
\ee
where
\be 
D=   (r_+ -2\gamma r_- - r_-)^2   [r_+^2 +\gamma(1+2\gamma)r_-^2 +(\gamma-1)r_+ r_-]^2  (r_+-\gamma r_-)^4  \ ,
\ee
and $N(M,q)$ is a rather complicated function that is always different from zero when $D(M,q)=0$. Curvature singularities exist if at least one of the following conditions is satisfied 
\bea 
r_+ -2\gamma r_- - r_- = 0 \ ,\label{sing1} \\
r_+^2 +\gamma(1+2\gamma)r_-^2 +(\gamma-1)r_+ r_- =0\ ,\label{sing2}  \\
r_+-\gamma r_- = 0 \ . \label{sing3}
\eea
The first singularity implies that the specific mass must be given by
\be
\frac{M^2}{q^2} = \frac{\eta^2 (1+3\gamma)^2 }{ 2(1+\gamma)(1+2\gamma)}\ ,
\ee
an expression which implies that
\be
R \rightarrow \infty \quad {\rm for} \quad \Bigg\{ \begin{array}{ll} 
                                                   \gamma<-1 \ , \gamma> -1/2 & \  {\rm if} \ \eta_2 = 1 \\
																									\gamma\in (-1,-1/2)  &\  {\rm if} \ \eta_2 = -1 
																									\end{array}
\ee
We see that, depending on the value of the parameter $\gamma$,  phase transitions occur for black holes with electric and phantom charges.

\begin{figure}
\includegraphics[scale=0.5]{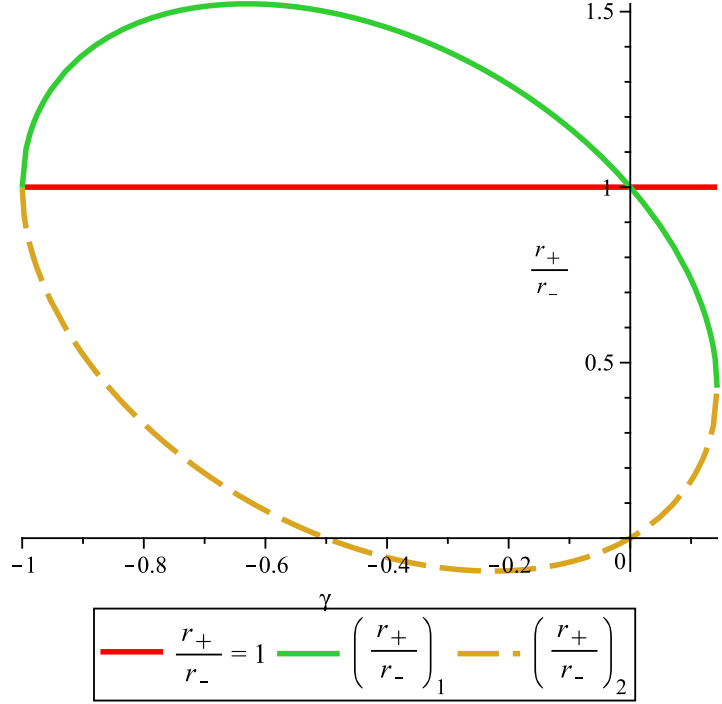}
\caption{The ratio $r_+/r_-$ as a function of the parameter $\gamma$ as follows from the singularity condition (\ref{sing2}). Since $r_- <r_+$, singularities can exist only the interval $\gamma\in (-1,0)$.} 
\label{fig1}
\end{figure}

For the second singularity condition (\ref{sing2}) we obtain two possible solutions
\be
\left(\frac{r_+}{r_-}\right)_{1,2}= \frac{1}{2} \left( 1-\gamma \pm \sqrt{1-6\gamma-7\gamma^2}\right) \ ,
\ee
which are real only within the interval $\gamma\in (-1,1/7)$. On the other hand,  for this interval of the parameter $\gamma$ we know from Eq.(\ref{gamma}) that $\eta_1=1$ and 
$0<r_-<r_+$. This implies that only in the case of pure dilatonic fields ($\eta_1=1)$ singularities are possible. In Fig. \ref{fig1}, we illustrate the behavior of
the ratio $r_+/r_-$ to find out the exact region where singularities are present. Indeed, since in this case $r_+/r_->1$, Fig. (\ref{fig1}) shows that only for
$\gamma \in (-1,0)$, solutions are allowed. Finally, from the condition (\ref{cond1}), one can see that only the case $\eta_2=1$ is allowed in the interval $0<r_-<r_+$.
We conclude that the second singularity condition cannot be satisfied in the case of phantom black holes ($\eta_1=\eta_2=-1)$.

Finally, from the third singularity condition (\ref{sing3}), we obtain that
\be 
r_+ = \gamma r_- \quad  {\rm i.e.}\quad  M^2 = \frac{2\eta_2 \gamma}{1+\gamma }  q^2 \ .
\label{cond3}
\ee
Since $r_+>r_-$, as stated in Sec. \ref{sec:pdbh}, the singularity exists only for $\gamma>1$ which, according to Eq.(\ref{gamma}), implies that $\eta_1=-1$
 and that $\eta_2=1$, in order for $M^2$ to be positive. Then, from the condition (\ref{cond1}) it follows that $\lambda_+ = 1-\lambda^2 >0$. This means that 
this singularity is present only in theories with $\lambda^2<1$ and phantom dilatonic field.

According to the formalism of GTD, the above singularities correspond to phase transitions of the corresponding black holes. 
On the other hand, according to classical black hole thermodynamics, the phase transition structure is determined by the behavior of the heat capacity. A more general 
structure is obtained by considering all the response functions of the system. In the case under consideration, the fundamental equation is given as the homogeneous function $S=S(M,q)$ which, in analogy to ordinary thermodynamics,  
 leads to the following heat capacity and response functions \cite{callen}
\be
C_q = T \left(\frac{\partial S}{\partial T}\right)_q\ , \quad \alpha_\phi =\frac{1}{q}  \left(\frac{\partial q}{\partial T}\right)_\phi\ , \quad
\beta_T = -\frac{1}{q} \left(\frac{\partial q}{\partial \phi} \right)_T\ ,
\ee
where for simplicity we denote as $\phi$ the intensive variable dual to the charge $q$. Using the fundamental equation (\ref{ent}), we obtain
($S_M = \partial S /\partial M$, etc.) 
\be
C_q = -\left(\frac{S_M^2}{S_{MM}}\right)_q = -\frac{2\pi r_+^{1+\gamma} (r_+ - r_-) ^{1-\gamma} (r_+ -\gamma r_-)}{r_+-2\gamma r_- - r_-}\ ,
\ee
\be
\alpha_\phi = - \frac{1}{q} \left(\frac{S_M^2}{S_{Mq}}\right)_\phi = - \frac{\pi}{\gamma q} \sqrt{\frac{2}{\eta_2(1+\gamma)} }\ 
\frac{ r_+ ^{3/2+\gamma} (r_+ - r_-)^{1-\gamma} (r_+ - \gamma r_-)} { r_-^{3/2}} \ ,
\ee
and
\be
\beta_T = \frac{1}{qT} \left(\frac{1}{S_{qq}}\right)_T = \frac{1}{4\pi qT} 
\frac{(r_+-r_-)^{1+\gamma} (r_+-\gamma r_-)}
{ r_-^\gamma[r_+^2 +\gamma(1+2\gamma)r_-^2 +(\gamma-1)r_+ r_- ] } \ .
\ee
The identification of the curvature singularities can now be performed as follows. The singularity (\ref{sing1}) coincides with a divergence of the heat capacity $C_q$ whereas the singularity (\ref{sing2}) corresponds to the blow up of the compressibility $\beta_T$. This implies that the singularities (\ref{sing1}) and (\ref{sing2}) determine second-order phase transitions. 

The third singularity (\ref{sing3}), however, is different. In fact, it does not correspond to a divergence of the heat capacity or the response functions; instead, it occurs at the point where  all of them vanish.  This means that at the  singularity  $r_+ = \gamma r_-$, the black hole undergoes a transition from a stable state to an unstable state  (or viceversa) which is, moreover, accompanied by a divergence of all second derivatives of the fundamental equation. In addition, the determinant of the thermodynamic metric (\ref{gd}) diverges and so the geometric description breaks down.

It is interesting to notice that in the limiting case $\gamma=1$, only one singularity is  present. Indeed, a straightforward computation shows that in this case the curvature scalar becomes 
\be R = 
{\frac {6\,r_-  ^{6}+57\, r_+  r_-  ^{5}
-r_+ ^{2} r_-  ^{4}+14  r_+ ^{3}  r_-  ^{3}
-8   r_+ ^{4}  r_- ^{2}-7  r_+ ^{5} r_-  +3\ r_+ ^{6
}}{ 2\pi^2 r_+^4 \left( 3 r_-  ^{2}+  r_+ ^{2} \right) ^{2} \left( 3   r_-  - r_+  \right) ^{2}   }}\ ,
\ee
indicating that only the first singularity survives which corresponds to the divergence of the heat capacity $C_q$.
Since $\gamma = (1-\eta_1\lambda^2)/(1+\eta_1\lambda^2)$, the limiting case $\gamma=1$ corresponds to $\lambda=0$, i.e., when the dilatonic and electromagnetic 
fields are minimally coupled in the action (\ref{action}). 
This shows that the presence of a non-minimal coupling in the action drastically affects the thermodynamic properties of black holes. 

For the case $\gamma=1$  it is possible to invert the fundamental equation (\ref{ent}) to obtain
\be
M=\frac{1}{2\sqrt{\pi S}}\left( S + \eta_2 \pi q^2\right) \ ,
\label{mass}
\ee
which determines the fundamental equation in the mass representation, $M=M(S,q)$, for which a GTD analysis can be performed \cite{ro12}. 
Indeed, the GTD approach was formulated in such a way that it can be applied to any representation. However, when performing concrete calculations, 
it is necessary to consider some details related 
to a change of a representation. If a fundamental equation is not invertible, there is only one possible representation and GTD allows us to carry out the complete analysis in that particular representation. If, on the contrary, a fundamental equation is invertible, there are (at least) two representations. 
On the phase space ${\cal T}$, a change of representation can be interpreted as a conformal transformation \cite{blnq14}. Consequently, for a geometric construction, like GTD, to be invariant with respect to changes of representation, it is necessary to demand conformal invariance. Then, the metric $G$ of ${\cal L}$ must be conformal and Legendre invariant. These conditions are very restrictive and leave us with practically no useful metrics for ${\cal T}$ \cite{blnq14}. 
In particular, the metric 
$G^{^{I/II}}$, as given in Eq.(\ref{gup}), is not conformal invariant and can, therefore, lead to  inconsistent results when applied to different representations. 
 This explains the inconsistencies found in \cite{ro12}. 

Nevertheless, there is simple solution to this problem, namely, it is always possible to consider a change of representation as a  coordinate transformation in ${\cal E}$. For concreteness, let us consider the above example with $\gamma=1$ for which the $S-$representation 
is determined by the fundamental equation (\ref{ent}) and the $M-$representation by Eq.(\ref{mass}).  The coordinates of ${\cal E}$ 
in the $S-$ representation are $\{M,q\}$. Let us introduce in ${\cal E}$ new coordinates $\{S, q'\}$ by means of the equations
\be 
M=\frac{1}{2\sqrt{\pi S}}\left( S + \eta_2 \pi q'^2\right) 
 \ , \quad q = q'\ ,
\ee
which is a well-defined coordinate transformation because its Jacobian is different from zero, $J= \partial M/\partial S \neq 0$. Obviously, such a coordinate transformation does not affect the geometric properties of the metric $g$ of ${\cal E}$ and, consequently, the thermodynamic properties of the corresponding 
system. This issue will be considered in more detail elsewhere \cite{quev16}.


\section{Final remarks}
\label{sec:con}

In this work, we used the formalism of GTD to analyze the thermodynamic properties of phantom dilatonic black holes. We considered a particular class of 
spherically symmetric black holes which is characterized by two parameters, namely, mass $M$ and electric charge $q$. The dilatonic field depends on $M$ and $q$ as 
well. The corresponding action contains two parameters, $\eta_1$ and $\eta_2$,  that determine the phantom nature of the electric charge and the 
dilatonic field. Phantom fields are characterized by negative energy densities at the level of the action.

We used the fundamental equation of this class of phantom black holes to construct a Legendre invariant metric for the corresponding equilibrium space. The investigation of the thermodynamic curvature shows that there are three different types of singularities which correspond to phase transitions. 
We also analyze the phase transition structure by using the standard methods of classical thermodynamics. We found that two phase transitions predicted by GTD correspond to divergences of the heat capacity and  compressibility, indicating that the results of GTD are compatible with classical black hole thermodynamics.
As for the third thermodynamic singularity, we found that it corresponds to a transition of the black hole in which it undergoes a drastic change of its stability properties. This type of transition occurs only in the presence of a phantom charge. Therefore, we interpret this transition as a consequence of the exotic nature
of the phantom black hole.  

Our results contribute to clarify some inconsistencies found by Rodrigues and Oporto in \cite{ro12}, when applying the formalism of GTD to the case of exotic black holes 
with phantom charges. In addition, we found a different type of phase transition related to a change between states with different stability properties. We explain this as a result of the exotic nature of the matter that generates the black hole. Another possible explanation could be that the Ehrenfest scheme, which is used in classical black hole thermodynamics to determine the phase transition structure, needs a generalization for the case of exotic matter. In the case of certain ordinary laboratory systems, it is already known that the Ehrenfest scheme fails to predict the observed phase transitions (see, for instance, \cite{jae98} and the references cited therein). Maybe we are now confronted with a similar situation  in black hole thermodynamics. We expect to analyze this task in a future investigation.

\section*{Acknowledgements}

This work was carried out within the scope of the project CIAS 1790
supported by the Vicerrector\'\i a de Investigaciones de la Universidad
Militar Nueva Granada - Vigencia 2015. This work was partially supported
by DGAPA-UNAM, Grant No. 113514, and CONACyT, Grant No. 166391. We thank an anonymous 
referee for many interesting comments and suggestions which led to an extensively revised
version of our original manuscript. 


\appendix

\section{Legendre invariance}
\label{sec:leginv}

In GTD, we use the terminology and some concepts  which are commonly used  in differential geometry and relativistic field theories, but
not in thermodynamic geometry. In this appendix, we explain such terms and conceptual issues. We will use the ideas of special 
relativity as an example, without pretending to be mathematically rigorous. More details can be found in the textbooks \cite{mtw,choq}.

Consider a  2-dimensional manifold $M^2$ endowed with the Minkowski metric 
\be
ds_2^2 = dt^2 - dx ^2 \ .
\label{mink2}
\ee   
Under the action of a Lorentz transformation $\{t,x\}\rightarrow \{\tilde t,\tilde x\}$ given by 
\be
t=\frac{1}{\sqrt{1-v^2}}\left(\tilde t + v \tilde x\right)\ ,\quad x = \frac{1}{\sqrt{1-v^2}}\left(\tilde x + v \tilde t\right)\ ,
\label{lor2}
\ee
where $v$ is a constant, the line element (\ref{mink2}) transforms into
 \be
ds_2^2 = d\tilde t^2 - d\tilde x ^2 \ .
\label{mink2n}
\ee   
We say then that the line element (\ref{mink2}) remains invariant under a Lorentz transformation or, equivalently, it is Lorentz
invariant or  it preserves Lorentz invariance. One can also say that the Minkowski metric is Lorentz invariant. The physical theories 
that are based upon the Minkowski metric (for instance, special relativity and gauge field theories) are called Lorentz invariant. 

The line element (\ref{mink2}) is
by definition a scalar, but it does not imply that it is invariant with respect to any arbitrary coordinate transformation. Consider, for instance, 
the transformation 
\be
t=\alpha_1 \tilde t \ , \quad x = \alpha_2 \tilde x + \alpha_3 \tilde t ^2 \ ,
\label{tra1}
\ee
where $\alpha_1$, $\alpha_2$ and $\alpha_3$ are constants. Then, the line element (\ref{mink2}) transforms into
\be
ds_2^2 = (\alpha_1^2 - 4 \alpha_3^2\tilde t^2)d\tilde t ^2 - 4 \alpha_2\alpha_3 \tilde t d\tilde x d\tilde t - \alpha_2^2 d\tilde x ^2 \ ,
\label{mink22}
\ee
which is clearly different from (\ref{mink2}). We then say that the Minkowski metric is not invariant with respect to the transformation (\ref{tra1}). Nevertheless, 
the fact that a line element is a scalar implies a very important property, namely, that the geometric (and physical) properties of the corresponding manifold do not depend on the choice of coordinates (in general relativity, this property is called covariance). One property which is very important for our purposes is the curvature. It can easily be shown that for the line elements (\ref{mink2}), (\ref{mink2n}) and (\ref{mink22})  all the components of the Riemann curvature tensor $R_{abcd}$ vanish identically, i.e., the vanishing of the curvature is a property which does not depend on the choice of coordinates. 

We note that from the point of view of pure Riemannian geometry the Lorentz transformation (\ref{lor2}) and the non-linear transformation (\ref{tra1}) belong to
the class of diffeomorphisms with respect to which the Minkowski line element (\ref{mink2}) behaves as a scalar. The fact that the Lorentz transformation preserves in addition the functional form of the line element (i.e, the Lorentz transformation is an isometry) is a complementary condition that is not required in pure Riemannian geometry. If one is interested in only the geometric behavior of the line element as a scalar, it is not necessary to consider Lorentz transformations as something special. On the other hand, if one is interested in the physical consequences of a transformation that leaves also the functional form of the Minkowski line element invariant, the Lorentz transformations are important to understand the canonical laws of spacetime. We will see below that in GTD we impose Legendre invariance as 
a complementary condition to take into account the properties of classical thermodynamics. In this connection, one can by analogy say that the Legendre transformations are to GTD what the Lorentz transformations are to special relativity.

Consider now the following coordinate transformation $\{t,x,y\}\rightarrow \{\tilde t,\tilde x, \tilde y\}$ with
\be
t=\frac{1}{\sqrt{1-v^2}}\left(\tilde t + v \tilde x\right)\ ,\quad x = \frac{1}{\sqrt{1-v^2}}\left(\tilde x + v \tilde t\right)\ .
\quad y =\frac{1}{\sqrt{1-v^2}}\left(\tilde y+ v \tilde t\right)\ ,
\label{tra2}
\ee
an let us  ask the question: Is the Minkowski line element (\ref{mink2}) invariant with respect to this transformation?  It is easy to see that the question 
is not well posed because the  Minkowski line element is 2-dimensional whereas the coordinate transformation (\ref{tra2}) involves three coordinates, i.e., it can 
be applied in a 3-dimensional manifold only in which the line element must contain an additional term, for instance,  
\be
ds_3^2 = dt^2 - dx ^2 - dy ^2 \ .
\label{mink3}
\ee   
This fact will be relevant below when considering Legendre transformations and Hessian metrics.

We now turn back to thermodynamics. It is well known that the laws of classical equilibrium thermodynamics can be written in different thermodynamic potentials, 
without affecting the properties of the systems under consideration \cite{callen}. This is to say that the properties of a thermodynamic system do not depend 
on the choice of the thermodynamic potential used to describe it. 
On the other hand, different thermodynamic potentials are always related by means of Legendre transformations \cite{callen}. 
Using the terminology explained above in this Appendix, we can say that equilibrium thermodynamics is Legendre invariant. One of the main goals of GTD is to incorporate this property into a geometric description of thermodynamics. To this end, it is necessary to define the Legendre transformations as coordinate transformations. This has been done long ago by Arnold \cite{arnold} who proved that any Legendre transformation can be represented as 
$\{Z^A\}\longrightarrow \{\widetilde{Z}^A\}=\{\tilde \Phi, \tilde E^a, \tilde I ^ a\}$ $(a=1,\ldots, n)$,   where the explicit relations between the old and the new coordinates are given in Eq.(\ref{leg}). Since a  Legendre transformation involves $2n+1$ coordinates, it must act on a $(2n+1)-$dimensional  manifold which is called thermodynamic phase space ${\cal T}$. Then, a metric defined on ${\cal T}$ by means of the line element
\be
ds^2_{2n+1} = G_{AB} dZ^A dZ^B 
\ee
is said to be Legendre invariant if the functional dependence of the components $G_{AB}$ does not change under a Legendre transformation. The Legendre invariance condition leads to a set of algebraic equations on the components $G_{AB}$ which for the case $n=2$ were given explicitly in \cite{quev07}.  By solving this set of equations, in GTD, we have found so far three classes of Legendre invariant metrics, namely, the two classes mentioned in Eq.(\ref{gup})
\be
\left(ds_{2n+1}^2\right)^{^{I/II}} \equiv G^{^{I/II}} = (d\Phi - \delta_{ab} I^a d E^b)^2 + (\xi_{ab} I^a E^b) (\chi_{cd} dI^c dE^d) \ ,
\label{gupap}
\ee
and a third class which is invariant under partial Legendre transformations
\be	
	\label{GIIIL}
	G^{^{III}}  =(d\Phi - I_a d E^a)^2  +  \left(E_a I_a \right)^{2k+1}  d E^a   d I^a \ , \quad I_a = \delta_{ab} I^b\ ,
\ee
where $k$ is an integer. 
To show the Legendre invariance, for instance, in the case of the metric (\ref{gupap}), it is sufficient to apply the Legendre transformation (\ref{leg}) to each of
the metric components. Then, we obtain
\be
G^{^{I/II}} = (d \tilde \Phi - \delta_{ab} \tilde I^a d \tilde E^b)^2 + (\xi_{ab} \tilde I^a \tilde E^b) (\chi_{cd} d\tilde I^c d\tilde E^d) \ ,
\ee
which shows clearly that the functional dependence of the metric components does not change, as demanded by Legendre invariance.  Thus, we have seen that the
phase space ${\cal T}$ can be considered in GTD as an auxiliary space which is necessary in order to handle correctly Legendre transformations as coordinate 
transformations. Furthermore, to study the properties of thermodynamic systems in GTD,  we consider the $n-$dimensional equilibrium space ${\cal E}$ with coordinates 
$\{E^a\}$  as a subspace of ${\cal T}$ 
defined by the embedding map $\varphi: {\cal E} \rightarrow {\cal T}$ such that the pullback $\varphi^*$ satisfies the condition 
$\varphi^*(d\Phi - \delta_{ab} I^a d E^b) = 0$, and induces on ${\cal E}$ the thermodynamic metric $g=\varphi^*(G)$. 
In the case of $G^{^{II}}$, the induced metric   $g^{^{II}}$
is explicitly given in Eq.(\ref{gdown}). Accordingly, we say that the metric $g$ of ${\cal E}$  is Legendre invariant, if it can be obtained 
from a Legendre invariant metric $G$ of ${\cal T}$ as $g=\varphi^*(G)$. 

Consider now  the class of Hessian metrics 
\be
ds_n^2 \equiv g^{^H} = \frac{\partial^2 H}{\partial E^a \partial E^b} dE^a d E^b \ , 
\label{hess}
\ee
where $H(E^a)$ is a thermodynamic potential and let us ask the question whether a Hessian metric is Legendre invariant. It is true that this line element is a scalar, but as explained above, it does not imply that it is invariant with respect to arbitrary coordinate transformations. In particular, only coordinate transformations of
the form $\{E^a\}\rightarrow \{\tilde E^a\}$ are allowed in the  $n-$dimensional manifold ${\cal H}$ defined by the Hessian metric (\ref{hess}). Since the Legendre transformations involve $2n+1$ coordinates, they cannot be applied in the $n-$dimensional manifold ${\cal H}$ and, therefore, we need to introduce a $(2n+1)-$dimensional manifold ${\cal T}^{^H}$ with metric $G^{^H}$ and coordinates $Z^A$ such that $\varphi^*(G^{^H}) = g^{^H}$. A straightforward computation shows that 
the metric
\be
G^{^{H}} = (d\Phi - \delta_{ab} I^a d E^b)^2 + \delta_{ab}  dE^a dI^b \ 
\label{ghess}
\ee
induces the Hessian metric (\ref{hess}), but it is not Legendre invariant. In this sense, we say that Hessian metrics do not preserve Legendre invariance.

Finally, let us consider the question about the uniqueness of the Legendre invariant metrics $G^{^{I/II}}$ and  $G^{^{III}}$ used in GTD. In all the cases, 
the first term contains the fundamental 1-form $\Theta =  d\Phi - \delta_{ab} I^a d E^b$ which, according to Darboux theorem, is defined modulo an
arbitrary  nonzero multiplicative function $f=f(Z^A)$, i.e., $\Theta$ and $f \, \Theta$ are equivalent. In this sense, the GTD-metrics of the phase space 
are not unique. However, 
this arbitrariness does not influence the thermodynamic metric $g=\varphi^*(G)$ because due to the property $\varphi^*(f\Theta) = f \varphi^*(\Theta)=0$, 
the function $f$ does not appear in the thermodynamic metric. Furthermore, the second term of the GTD-metrics can also be multiplied by a nonzero Legendre
invariant function $\Lambda = \Lambda (Z ^A)$. However, if we impose the additional physical condition that the curvature of the equilibrium space vanishes 
in the case of a thermodynamic system with no thermodynamic interaction (more details will be given in Appendix \ref{sec:tdint}), then $\Lambda$ is reduced
to a constant which can be set equal to one, without loss of generality. In this sense, all the metrics we use in GTD for the equilibrium space are unique. 
This is shown explicitly in the case of the final expression for the metric $g^{^{II}}$, given in Eq.(\ref{gdown}), which we use in GTD to describe 
black holes.

\section{Curvature and thermodynamic interaction}
\label{sec:tdint}

The idea of using curvature as a measure of interaction is due originally to A. Einstein. In general relativity, the curvature of the 4-dimensional spacetime manifold represents the gravitational interaction. If no gravitational interaction exists, the spacetime is flat. The same idea has been shown to be true also in the case of the  electromagnetic, weak and strong interactions (gauge interactions) which are determined by the curvature of a different higher-dimensional manifold (a principal fiber bundle). This means that all the fundamental interactions known in Nature have a geometric description in which the curvature is a measure of the interaction
(see, for instance, \cite{frankel}).  By analogy, one of the goals of GTD is to propose a geometric description of thermodynamics in which the corresponding 
curvature represents the effective thermodynamic interaction. To this end, we use Legendre invariance as a guidance principle. This idea is based again on the 
known properties of field theories. Indeed, to construct general relativity, Einstein used the diffeomorphism invariance whereas gauge field theories are based upon the gauge invariance. In each case, the invariance correspond to transformations which leave invariant the properties of the underlying theory.
Accordingly, in GTD we propose to use Legendre invariance as a guidance principle because equilibrium thermodynamics is Legendre invariant. This is the geometric intuitive approach we have been using in GTD to interpret the thermodynamic as representing the
thermodynamic interaction. This approach has been presented with some detail in \cite{qstv10}. A more physical approach based upon 
the interpretation of the thermodynamic metrics used in GTD as the second moment of the fluctuations of a new thermodynamic potential will be mentioned in Appendix 
\ref{sec:phys}.

As mentioned in Appendix \ref{sec:leginv}, the GTD-metrics were obtained by applying the condition of Legendre invariance in a mathematically consistent way, i.e., by 
introducing the thermodynamic phase space ${\cal T}$. However, in order to take into account the above condition that the curvature of the equilibrium space represents the thermodynamic interaction, we must demand that in the particular case where no interaction is present (ideal gas), the curvature vanishes. This is therefore an additional physical condition that must be imposed on the GTD-metrics and, in fact, a straightforward computation shows that this is true for the metrics presented in Appendix \ref{sec:leginv}. This means that the equilibrium space of the ideal gas must be flat; a detailed analysis of this case was performed in \cite{qsv15}. 

Thus, we see that the GTD-metrics are not only Legendre invariant, but they also satisfy the physical condition of leading to flat equilibrium spaces when thermodynamic interaction is lacking. In fact, these two conditions were used in the original formulation of GTD  \cite{quev07} to select viable metrics. More recently, by using a group theoretical approach based upon infinitesimal Lie symmetries, the most general metric (with $n=2$) was obtained which is invariant under the action of infinitesimal Legendre transformations \cite{gl14}. However, this general metric does not satisfy the physical condition of leading to a flat equilibrium manifold in 
the case of the ideal gas, i.e., it cannot be used in GTD. To show this,  let us consider the general metric obtained in Eq.(33) of Ref. \cite{gl14} which, using  
the conventions and notations of the present work, can be written as
\be
G^{inf} = \Theta ^ 2 + 2 \Omega (dE^1 d I^2 - dE^2 dI^1) \ ,
\ee
where  the function $\Omega=\Omega(E^1,E^2,I^1,I^2)$ is nonzero and invariant under infinitesimal Legendre transformations. The induced metric of the equilibrium 
space reads
\be
g^{inf}= \varphi^*(G^{inf}) = 2 \Omega \left\{ \Phi_{12} [ (dE^1)^2 - (dE^2)^2] + (\Phi_{22}- \Phi_{11}) dE ^1 d E^2 \right\}  
\ , 
\ee
with 
\be
\Phi_{12} =  \frac{\partial^2 \Phi}{\partial E^1 \partial E^2}, \quad {\rm etc.} 
\ee
Here we have used the condition $\varphi^*(\Theta) = \varphi^*(d\Phi - I^1 dE^1 - I^2 dE^2)=0$. If we now consider the fundamental equation for an ideal gas, which 
in the entropy representation is essentially $\Phi = S= \ln U  + \ln V + $ const. with $U=E^1$ and $V=E^2$,  it can be shown that there is no function $\Omega$ 
that leads to a zero curvature tensor for ${\cal E}$. In other words, the equilibrium space constructed from the metric $G^{inf}$ for an ideal gas is not flat; therefore, $G^{inf}$ cannot be used in GTD. In fact, the GTD-metrics $G^{^{I/II}}$ and $G^{^{III}}$ are by no means related to $G^{inf}$. 

Finally, let us notice that the results presented in Ref. \cite{gl14} support the results obtained in GTD. Indeed, the main result of Ref. \cite{gl14} is that the
most general metric which is invariant under infinitesimal Legendre transformations does not lead to a flat equilibrium space for the ideal gas. Therefore, to reach this physical goal, it is necessary to use non-infinitesimal (discrete) Legendre transformations, and this is exactly what we have been doing in GTD.

\section{On the physical significance of the GTD-metrics}
\label{sec:phys}

In this appendix, we investigate the question about the physical significance of the metrics obtained in GTD under the condition of Legendre invariance 
and that the thermodynamic curvature is a measure the thermodynamic interaction. To this end, we will consider the main conceptual ideas of 
classical thermodynamic fluctuation theory \cite{ll77}. Suppose the equilibrium state of a thermodynamic system is determined by the fundamental equation
$H(E^a)$. Let us denote by $dE^a$ the infinitesimal deviations of the variables $E^a$ from the equilibrium state. In very broad terms, in fluctuation theory one considers the deviations of $H$ by means of the expansion 
\be
H(E^a+dE^a) = H(E^a) + \frac{\partial H}{\partial E^a} d E^a + \frac{1}{2} \frac{\partial^2 H}{\partial E^a\partial E^b } dE^a d E ^b + \cdots 
\ee
If we choose $H$ as the 
total entropy of the Universe $S$ and recall that it reaches a maximum at equilibrium, i.e., $\partial S/\partial E^a=0$, then the second moment of the fluctuation is essentially given by the components of the Ruppeiner metric \cite{rup95}. This important result provides Ruppeiner metric  with a clear physical significance, and permits to find the connection with information geometry. It is also  important to note that in this case the coordinates must correspond to conserved quantities in order for the first derivative to vanish and for the Hessian to determine the components of the thermodynamic metric. This is, of course, not always the case. For 
instance, if we have a look at the thermodynamic metrics $g$, induced by the Legendre invariant metrics $G$, in the 
entropy representation, it is easy to see that their components do not correspond to the second moment of the entropy fluctuations. Nevertheless, we will now briefly show that it is possible to introduce new coordinates in the phase space such that the thermodynamic metrics can be interpreted in terms of the second  moment of the fluctuations of a different thermodynamic potential.

Let us first notice that all the GTD-metrics given in Eqs.(\ref{gupap}) and (\ref{GIIIL}) can be rewritten as 
\be
G = (d\Phi - \delta_{ab}I^a dE^b)^2 + h_{ab} dE^a dI ^b \ ,
\label{gtdmet}
\ee
where the components $h_{ab}$ are functions of the coordinates $E^a$ and $I^a$. If we calculate the thermodynamic metric $g=\varphi^*(G)$ under the condition that the canonical contact form $\Theta$ vanishes, 
$\varphi^*(\Theta)= \varphi^*(d\Phi - \delta_{ab}I^a dE^b) = 0$, we obtain $g= h_{ab}\delta^{cb} \Phi_{,cd} dE^a dE ^d$ whose components, as mentioned above, cannot be interpreted 
in the framework of fluctuation theory. However, let us recall that the metric $G$ is fixed in the coordinates $Z^A=\{\Phi, E^a, I^a\}$ as a result of imposing Legendre invariance, but we can still perform a coordinate transformation of the form $Z^A\rightarrow \bar Z ^{A} = \{F, X^a, Y^a\}$, where in general
\be
F=F(\Phi, E^b, I^b)\ , \quad X^a= X^a (\Phi, E^b, I^b)\ , \quad Y^a=Y^a(\Phi, E^b, I^b)\ .
\ee
The only condition to be imposed is that the Jacobian of the transformation is different from zero. This procedure is similar to the one we mentioned for the Minkowski metric in Appendix \ref{sec:leginv}, namely, once the Minkowski metric is fixed as (\ref{mink2}) in order to be Lorentz invariant, we can perform any coordinate transformation, for instance (\ref{tra1}), without changing the geometric properties of the Minkowski spacetime. 

Applying the above coordinate transformation to the second term of $G$, i.e.,  $h_{ab}dE^adI^b$,  it is easy to see that by choosing the new coordinates 
$X^a= X^a (E^b, I^b)$, and $Y^a=Y^a( E^b, I^b)$ in the appropriate way, the second term can always be brought to the form  $\delta_{ab}dX^adY^b$. Now let us
consider the 1-form $\bar \Theta = f(dF - Y_a d X^a)$, where $f=f(F,X^a,Y^a)$ is a nonvanishing function. Indeed, according to the Darboux theorem, this is canonical 
contact form in the new coordinates $\bar Z ^A$.  Now, let us ask the question whether the coordinate transformation $Z^A\rightarrow \bar Z ^A$ can be used also 
to identify the two canonical contact forms, i.e., if the differential equation $d\Phi - I_a dE^a = f(dF - Y_a dX^a)$ holds\footnote{In more technical terms, we are simply asking whether the transformation  $Z^A\rightarrow \bar Z ^A$ is a contactomorfism.}. To answer this question, it is necessary to compute the corresponding
integrability conditions. Lengthy calculations show that they are not satisfied in general for any of the GTD-metrics. Nevertheless, a detailed study of the analytic form of the integrability conditions show that they are satisfied if we impose a ``deformation'' of the contact form, i.e. 
$\bar \Theta \rightarrow f_0dF - f_a Y_a d X^ a$, where the nonvanishing  functions $f_0$ and $f_a$ can depend on all coordinates $F$, $X^a$ and $Y^a$. 
Summarizing, we have proved that it is always possible to find a coordinate transformation $Z^A\rightarrow \bar Z ^A$ that brings the GTD line element 
(\ref{gtdmet}) into the form
\be
\bar G = (f_0 dF - f_a Y_a dX^a)^2 + \delta_{ab} dX^a dY ^b \ .
\ee
This is then the line element of the phase space ${\cal T}$  in the new coordinates $\bar Z ^A$. Let us consider the corresponding equilibrium space ${\cal E}$ by means of the 
embedding map $\bar \varphi:  {\cal E} \rightarrow {\cal T}$ which is defined by the condition $\bar \varphi^*(dF - Y_a d X^a) = 0$. Then, the thermodynamic 
metric $\bar g= \bar \varphi^*(\bar G)$ induced in ${\cal E}$ can be expressed as
\be
\bar g = (f_0-f_a)(f_0-f_b) \frac{\partial F}{\partial X^a} \frac{\partial F}{\partial X^b } dE^a dE ^b
+ \frac{\partial^2 F}{\partial X^a\partial X^a} dE^a dE ^b \ .
\ee 
We see that this metric contains first  and second-order  derivatives of the new coordinate $F(X^a)$. Moreover, it depends on $n+1$ arbitrary functions $f_0$ and $f_a$. 
Once we specify a function $F(X^a)$, the metric can be calculated explicitly. This means that $F(X^a)$ plays the role of fundamental equation in the new 
coordinates. Let us also demand that $F(X^a)$ reaches an extremum value at equilibrium, i.e., 
$\partial F/\partial X^a=0$. Then, at each point of the equilibrium space the metric $\bar g$ reduces to 
\be
\bar g =    \frac{\partial^2 F}{\partial X^a\partial X^b} dE^a dE ^b \ .
\ee
If we now consider the infinitesimal fluctuations $dX^a$ of the potential $F$ around an equilibrium state $E^a$, we obtain
\be
F(X^a+dX^a) = F(X^a) + \frac{1}{2}   \frac{\partial^2 F}{\partial X^a\partial X^b} dE^a dE ^b \ .
\ee
We conclude that the components of the GTD-metrics in the equilibrium space can be interpreted as the second moment of the fluctuation of the new thermodynamic
potential $F$.

The above analysis shows that it is possible to relate the GTD-metrics with fluctuation theory.  Moreover, we have shown that GTD allows us to construct new thermodynamic potentials other than those that are usually constructed in classical thermodynamics by means of Legendre transformations. The physical significance 
of the new potential $F$, however, is not clear. We believe that it is necessary to construct new potentials for specific thermodynamic systems in order to 
investigate their physical meaning. 
In this brief comment, we only presented a brief scheme of the mathematical proof of the existence of new thermodynamic potentials. We will present a 
more detailed analysis elsewhere \cite{pqv16}. 

Probably, this result will also  allow us to interpret the thermodynamic curvature as a measure of the thermodynamic interaction by using a more physical approach. Indeed, the Ruppeiner scalar curvature can be related with the fluctuating structure size because the metric is the Hessian of a particular thermodynamic potential, namely, the entropy.  We can therefore expect a similar interpretation for the scalar curvature of the GTD-metrics, but now in terms of the new thermodynamic 
potential $F$.  In other words, the Ruppeiner metric establishes the connection between curvature and interaction by using
the entropy as the thermodynamic potential  whereas the GTD-metrics use quite different potentials.

\end{document}